\documentclass{jpsj-suppl}
\usepackage{txfonts} 

\title{Role of Three-Nucleon Forces in Neutron-Rich Nuclei Beyond $^{132}$Sn}

\author{L. \textsc{Coraggio}$^{1}$, A. \textsc{Gargano}$^{1}$, and N. \textsc{Itaco}$^{1,2}$}

\inst{$^1$INFN, Complesso Universitario di Monte S. Angelo, via Cintia,
  80126 Napoli, Italy\\
  $^2$Dipartimento di Fisica, Universit\`a di Napoli ``Federico II'',
  Complesso Universitario di Monte S. Angelo, via Cintia, 80126 Napoli, Italy\\}

\email{coraggio@na.infn.it}

\recdate{September 11, 2014}

\abst{The role of three-nucleon forces (3NF) in the description of nuclear
  structure properties is nowadays a main topic in the field of
  microscopic many-nucleon calculations.
We investigate the relative weight between effective two- and
three-nucleon forces in neutron-rich nuclei beyond the doubly-closed
$^{132}$Sn core within the realistic shell-model framework, studying
the evolution of the spectroscopic properties of $N=82$ isotones and
heavy tin isotopes.
This problem is tackled indirectly without explicitly taking into
account effective 3NF through the comparison of the results of
shell-model calculations obtained from realistic on-shell-equivalent
low-momentum potentials.}

\kword{Nuclear shell model, realistic nucleon-nucleon potentials,
  effective interactions}

\begin{document}
\maketitle

\section{Introduction}
The study of the effects of 3NF is currently a relevant issue in
nuclear structure calculations for many-body systems
\cite{Navratil07,Hagen07,Hagen12,Soma14}.
The renormalization of a high-momentum realistic high-precision
nucleon-nucleon ($NN$) potential via the $V_{\rm  low-k}$
\cite{Bogner01,Bogner02} or the similarity-renormalization group (SRG)
\cite{Bogner10} approaches provides on-shell-equivalent $NN$
interactions which are suitable for perturbative many-body
calculations, and that are characterized by a cutoff parameter $\Lambda$.
Actually, starting from these on-shell equivalent $NN$ potentials the
results of calculations for few-body and infinite systems are
dependent from the chosen cutoff owing to their different off-shell
behavior, and the cutoff independence can be restored introducing
many-body forces aside the $NN$ one \cite{Nogga04,Bogner05}.

As a matter of fact, in the effective field theory (EFT) framework -
where both $V_{\rm low-k}$ and SRG approaches are
grounded - the calculated physical observables are ideally independent
of the cutoff scale $\Lambda$, provided that nuclear two- and
many-body forces are taken into account \cite{Kol99}.
For nuclear potentials this can be done by way of the chiral 
perturbation theory (ChPT), which introduces nuclear two-, three-,
four-, ... body interactions on an equal footing \cite{Wei92,Kol94} able to
reproduce accurately the nucleon-nucleon ($NN$) data \cite{ME11}.

The regulator dependence of chiral potentials in many-body systems
has been recently investigated in both neutron and nuclear infinite
matter \cite{Coraggio13,Coraggio14}, where the $NNN$ potential has
been explicitly taken into account.
In particular, in \cite{Coraggio13} it has been evidenced the
crucial role of the $NNN$ potential in order to provide the cutoff
independence in infinite neutron matter.

The computational problem of including a 3NF in finite-nuclei
calculations is far more complicated than in infinite nuclear systems,
so in present paper we have faced this problem indirectly without
explicitly taking into account these forces.

We have studied the relative weight between effective two- and
three-nucleon forces in neutron-rich nuclei beyond the doubly-closed
$^{132}$Sn core within the realistic shell-model framework. 

To this end, starting from the high-precision $NN$ potential CD-Bonn
\cite{Machleidt01}, we have derived two on-shell-equivalent
low-momentum potentials by way of the $V_{\rm low-k}$ approach
\cite{Bogner02}.
These two potentials, which reproduce exactly the same
two-nucleon-system data of the original CD-Bonn potential, differ
because they are defined up to two cutoffs $\Lambda=2.1$ and 2.6
fm$^{-1}$.

As mentioned before, these potentials do not yield identical results in
the many-body problems, as testified by the different values of the
triton and $^{4}$He binding energies, and the differences should be
eliminated taking into account the 3NF corresponding to the chosen
cutoff and derived so as to reproduce the original-potential value of
the triton binding energy \cite{Nogga04,Bogner05}.

Starting from these two $V_{\rm low-k}$s, we have derived the
effective shell-model hamiltonians within the framework
of the time-dependent degenerate linked-diagram perturbation theory
\cite{Kuo71}. 
This approach provides both the single-particle (SP) energies and the
two-body matrix elements (TBME) of the residual potential without
resorting to any parameter fitted to experimental data \cite{Coraggio12}.

We have performed then shell-model calculations for nuclei outside the
doubly-closed $^{132}$Sn, focusing our attention on the evolution of
the spectroscopic properties of the $N=82$ isotones and heavy-mass tin
isotopes when adding valence nucleons.
This region is currently the subject of great experimental and
theoretical interest, especially in view of the production of new
neutron-rich nuclear species at the next generation of radioactive ion
beam facilities, in order to study the shell evolution versus the
number of the valence nucleons.

In the following section, we outline the perturbative approach to the
derivation of a realistic shell-model hamiltonian.
In Section 3, we present the results of our shell-model calculations,
performed by using the Oslo shell-model code \cite{EngelandSMC}, and
compare them with the experimental data to infer the role of the
missing 3NFs.
A short summary is given in the last section.

\section{Outline of calculations}
As mentioned in the Introduction, we derive the shell-model effective
hamiltonians within the framework of the time-dependent degenerate
linked-diagram expansion \cite{Coraggio09}.

To this end, an auxiliary one-body potential $U$ is introduced to
write down the hamiltonian as the sum of an unperturbed term $H_0$, 
which describes the independent motion of the nucleons, and a residual 
interaction $H_1$:

\[
H=\sum_{i=1}^{A} \frac{p_i^2}{2m} + \sum_{i<j} V^{ij}_{NN} = T + V_{NN} =
(T+U)+(V_{NN}-U)= H_{0}+H_{1}~~,
\]

\noindent
where $V_{NN}$ represents the input $NN$ potential.

The effective hamiltonian $H_{\rm eff}$ is obtained by way of the
Kuo-Lee-Ratcliff folded-diagram expansion in terms of the vertex
function $\hat{Q}$-box, which in the perturbative approach is built by
a collection of irreducible valence-linked diagrams \cite{Kuo71}.
We calculate the $\hat{Q}$-box including one- and
two-body Goldstone diagrams through third order in $H_1$
\cite{Coraggio12}.
Calculations beyond the third order in perturbation theory are
computationally prohibitive, so we have calculated the Pad\'e approximant
$[2|1]$ \cite{Ayoub79} of the $\hat{Q}$-box to obtain a
value to which the perturbation series should converge, as suggested
in \cite{Hoffmann76}.
The folded-diagram series is then summed up to all orders using the
Lee-Suzuki iteration method \cite{Suzuki80}.

The $\hat{Q}$-box contains one-body contributions, whose collection is the
so-called $\hat{S}$-box \cite{Shurpin83}.
The folded-diagram expansion of the $\hat{S}$-box represents the
theoretical SP energies, that will be employed in the shell-model
calculations.
The TBME will be then obtained by way of a subtraction procedure of
these SP energies from $H_{\rm  eff}$.

As mentioned in the Introduction, we employ as $V_{NN}$ a low-momentum potential 
$V_{\rm low-k}$ defined within a cutoff momentum $\Lambda$ by way of a
similarity transformation \cite{Bogner02}. 
This is a smooth potential which preserves exactly the on-shell
properties of the original $V_{NN}$ and is suitable for being used 
directly in nuclear structure calculations \cite{Coraggio09}.

We have derived from the high-precision CD-Bonn $NN$ potential
\cite{Machleidt01} two $V_{\rm low-k}$s corresponding to
$\Lambda=2.1$ and 2.6 fm$^{-1}$.

\begin{figure}[h]
\begin{center}
\includegraphics[scale=0.35,angle=0]{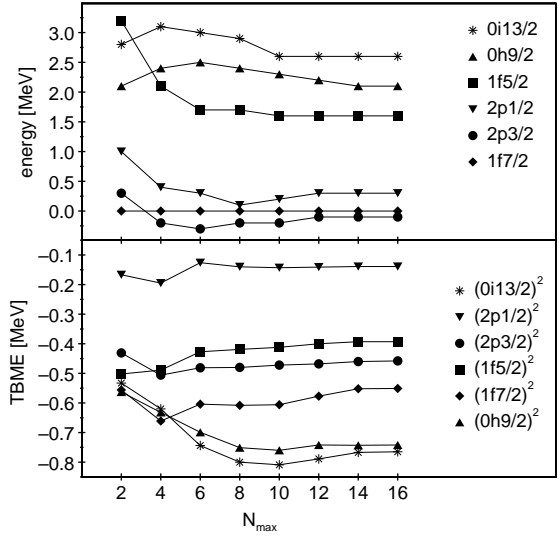}
\caption{Theoretical single-neutron relative energies (up) and neutron-neutron
  diagonal $J^{\pi}=0^+$ TBME (down) as a function of $N_{\rm max}$ (see text
  for details).}
\label{conv}
\end{center}
\end{figure}

It is worth to spend a few lines about the convergence properties of
the theoretical SP energy spectra and TBME as a function of the
dimension of the intermediate-state space.
In Fig. \ref{conv}, we report the single-neutron relative energy
spectrum and the diagonal $J^{\pi}=0^+$ TBME of the tin isotopes
calculated with the cutoff $\Lambda=2.6$ fm$^{-1}$ as a function of
the maximum allowed excitation energy of the intermediate states
expressed in terms of the oscillator quanta $N_{\rm max}$.
We show results with $\Lambda=2.6$ fm$^{-1}$ because the larger is
the cutoff the slower is the convergence rate.
From Fig. \ref{conv} it is clear that our results have practically
achieved convergence at $N_{\rm max}=16$.

\section{Results and comparison with experiment}
\begin{figure}[h]
\begin{center}
\includegraphics[scale=0.4,angle=0]{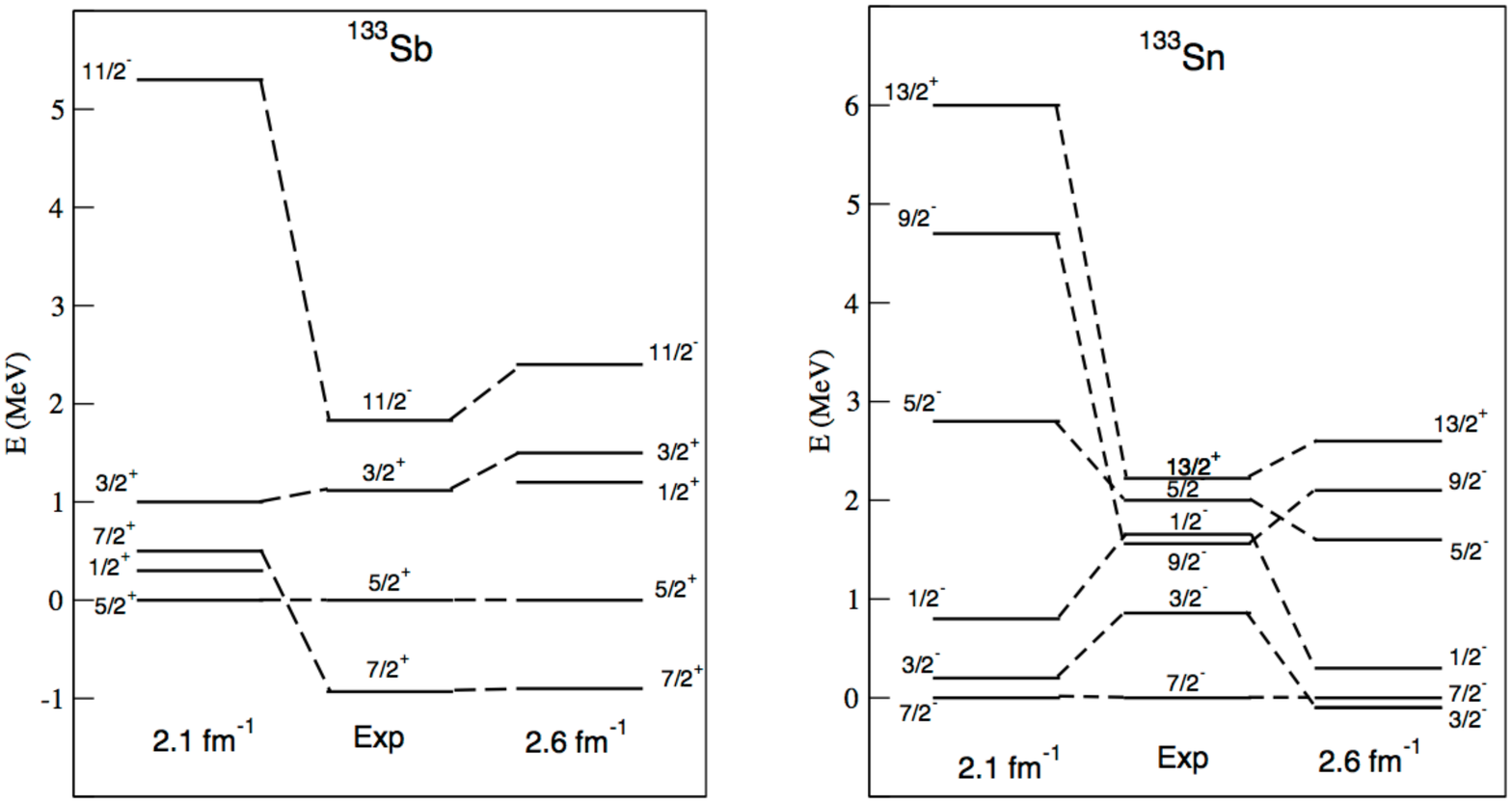}
\end{center}
\caption{Calculated and experimental single-particle spectra of
 $^{133}$Sb and $^{133}$Sb. The spectra are normalized with respect to
 $J=5/2^+$ and $J=7/2^-$ levels, respectively.}
\label{133sb}
\end{figure}

In Fig. \ref{133sb} the theoretical SP spectra of $^{133}$Sb and
$^{133}$Sn are compared with the experimental ones \cite{ensdf}.
They are normalized with respect to the proton $0d_{5/2}$ state and
neutron $1f_{7/2}$, respectively, in order to evidence the changes in
the spacings with the cutoff $\Lambda$.
It has to be pointed out that there is no experimental counterpart for
the proton $2s_{1/2}$ orbital and that the neutron $0i_{13/2}$ energy in
\cite{ensdf} is estimated as in \cite{Urban99}  from the observed
$10^+$ state at 2434 keV in $^{134}$Sb.

From the inspection of Fig. \ref{133sb}, it can be seen that the
calculated spin-orbit splitting between $1d_{3/2}$ and $1d_{5/2}$
levels is scarcely dependent on $\Lambda$, the experimental
separation being reasonably well reproduced. 
 
The situation is quite different for the calculated relative energies
of the $0g_{7/2}$ and $0h_{11/2}$ levels, whose spin-orbit partners
are outside the chosen model space.
In this case, the theoretical values show a strong dependence on the
cutoff $\Lambda$.
The most striking feature is that for $\Lambda=2.1$ fm$^{-1}$ the
discrepancies with respect to the experimental data are quite large,
while a better agreement is obtained with $\Lambda=2.6$ fm$^{-1}$ . 
In particular, it should be pointed out that with $\Lambda=2.1$
fm$^{-1}$ the $0h_{11/2}$ level lies far away from the other levels,
thus implying an enhanced shell closure at $Z=70$.
A much more reasonable spectrum of $^{133}$Sb is obtained when using
$\Lambda=2.6$ fm$^{-1}$  with the intruder $0h_{11/2}$ proton
state joining the 50-82 shell. 

The inspection of Fig. \ref{133sb} shows that similar conclusions can
be drawn for the single-neutron energies.
In fact, the $0h_{9/2}$, $0i_{31/2}$ relative energies are quite
sensitive to the choice of the cutoff, and a better agreement with
experiment is obtained with $\Lambda=2.6$ fm$^{-1}$.

\begin{figure}[ht]
\begin{center}
\includegraphics[scale=0.4,angle=0]{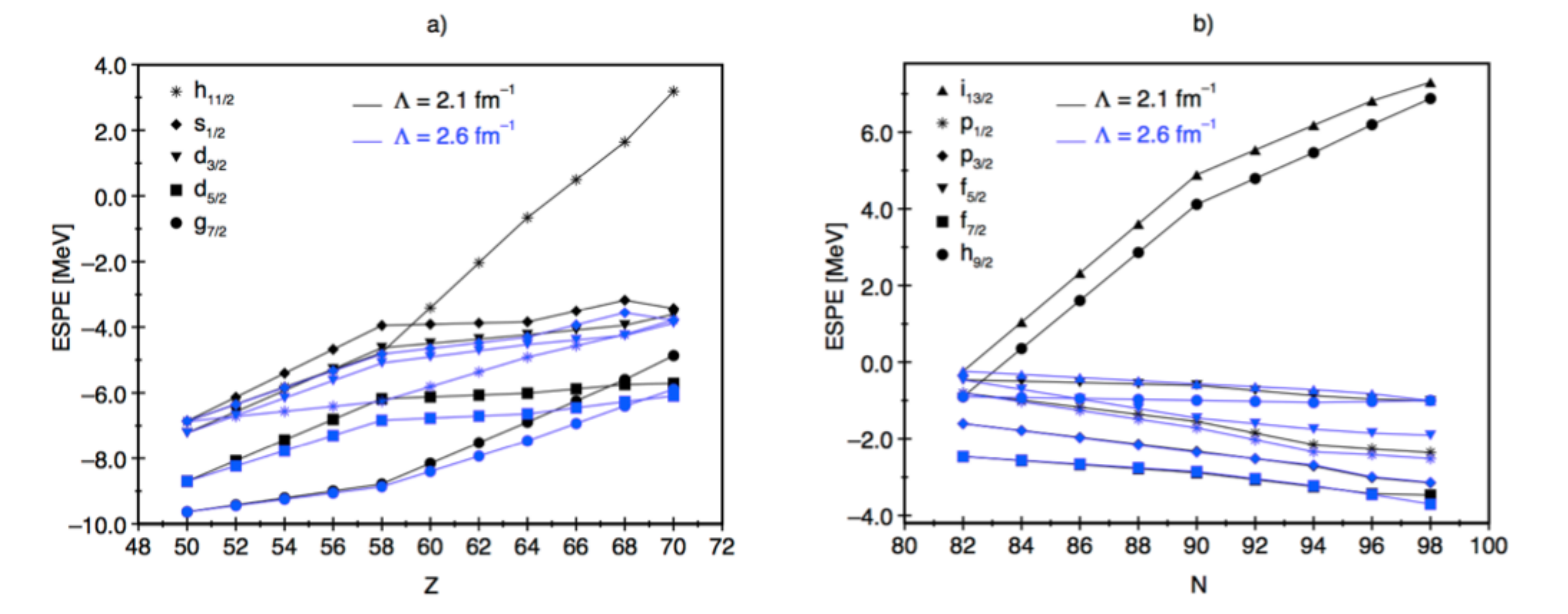}
\end{center}
\caption{Calculated effective single-particle energies for the $N=82$
  isotones (a) and heavy tin isotopes (b). The continuous black line refers to
  calculations performed with $\Lambda=2.1$ fm$^{-1}$, the blue one to
  $\Lambda=2.6$ fm$^{-1}$.}
\label{espe}
\end{figure}

The sensitivity to the cutoff value shown by the SP relative energies
of the orbitals with the spin-orbit partner outside the model space is
found also in the monopole component of the TBME.
In Fig. \ref{espe} we report the proton and neutron
effective single-particle energies (ESPE) calculated with both
$\Lambda=2.1$ and 2.6 fm$^{-1}$ as a function of $Z$ and $N$,
respectively.
We have employed the experimental SP energies available from the
$^{133}$Sb and $^{133}$Sn data \cite{ensdf}, in order to evidence the
role of the cutoff on the monopole term.
As for the proton $2s_{1/2}$ orbital, the SP energy has been
taken as the empirical value in \cite{Coraggio09}.

It can be seen that the $\pi 0h_{11/2},~\nu 0h_{9/2}$, and $\nu
0i_{13/2}$ ESPE exhibit a strong repulsive behavior when employing
$\Lambda=2.1$ fm$^{-1}$, while for the $\pi 0g_{7/2}$ ESPE this is far
less marked.
In particular, according to $\Lambda=2.1$ fm$^{-1}$ calculations it
should appear a strong shell closure at $Z=70$ and at $N=102$.

\begin{figure}[ht]
\begin{center}
\includegraphics[scale=0.25,angle=0]{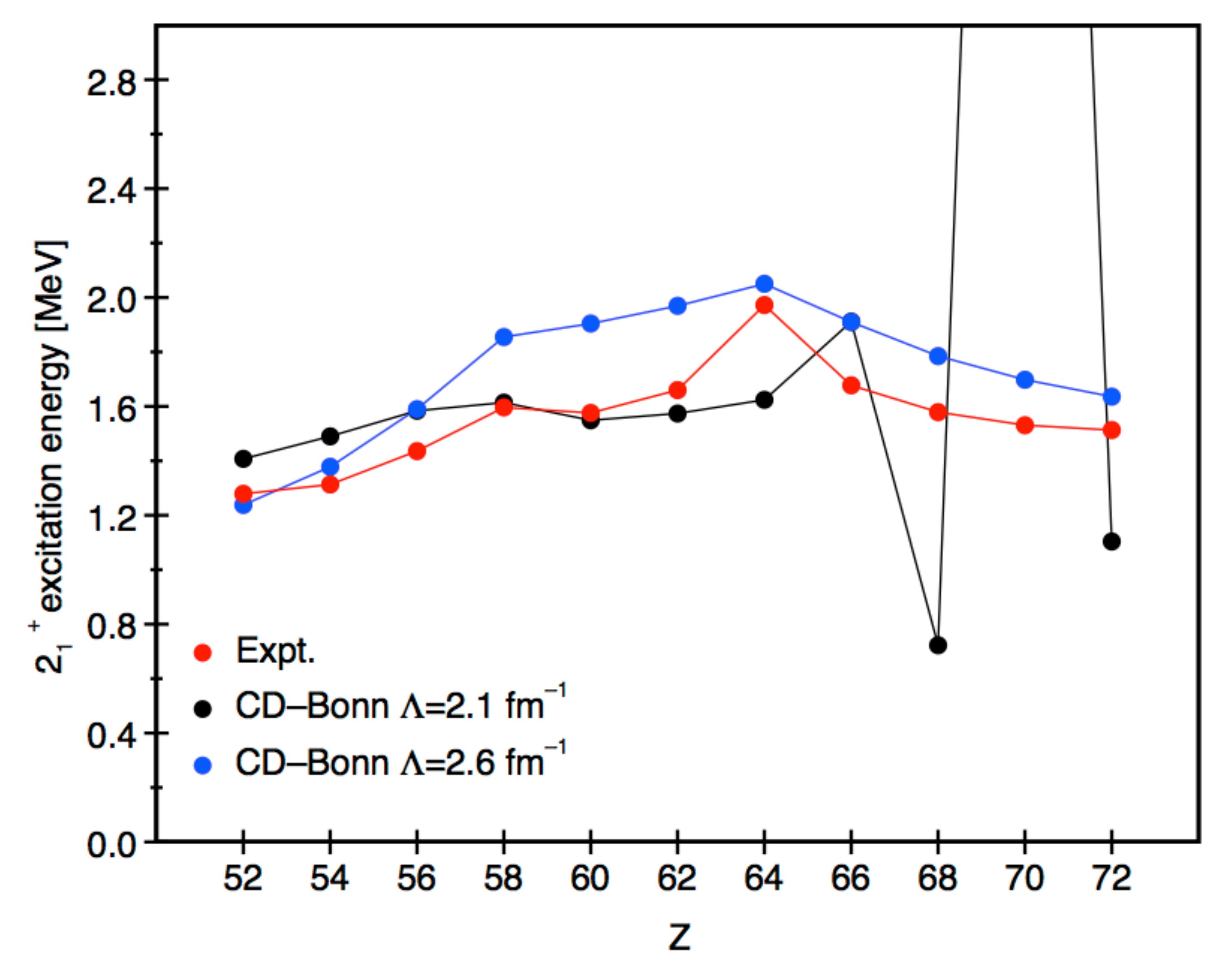}
\end{center}
\caption{Calculated and experimental excitations energies of the yrast
  $J=2^+$ states for the $N=82$ isotones. The continuous black line
  refers to calculations performed with $\Lambda=2.1$ fm$^{-1}$, the
  blue one to $\Lambda=2.6$ fm$^{-1}$.}
\label{N82J2p}
\end{figure}

\begin{figure}[ht]
\begin{center}
\includegraphics[scale=0.25,angle=0]{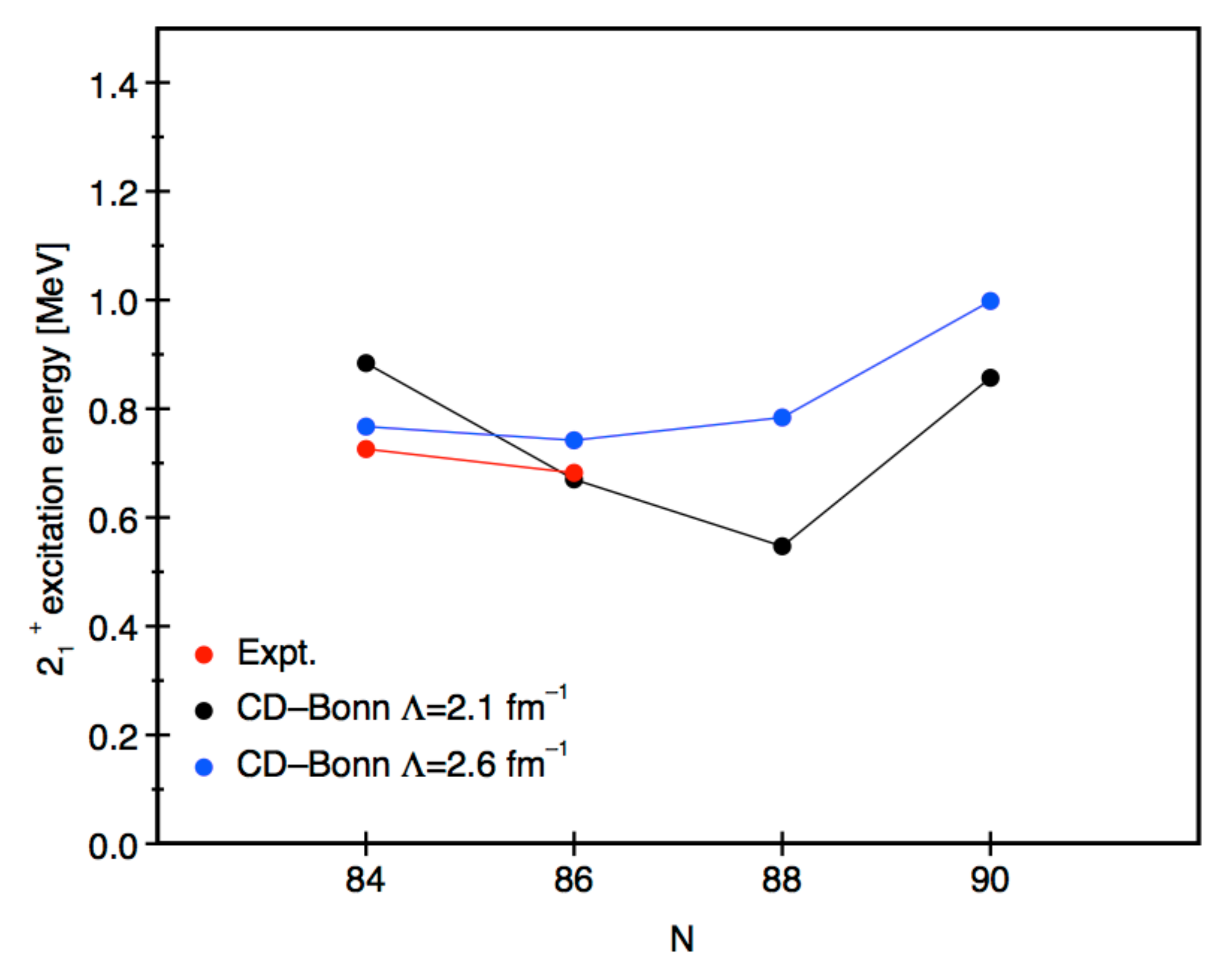}
\end{center}
\caption{Same as in Fig. \ref{N82J2p}, but for the heavy tin
  isotopes.}
\label{tinJ2p}
\end{figure}

In Fig. \ref{N82J2p} the behavior of the calculated and experimental
excitation energies of the yrast $J=2^+$ states for the $N=82$
isotones is reported.
Experimental data evidence a subshell closure
at $Z=64$ for $^{146}$Gd and no shell closure at $Z=70$ for
$^{152}$Yb, correctly reproduced with the cutoff $\Lambda=2.6$
fm$^{-1}$.
On the other side $\Lambda=2.1$ fm$^{-1}$ results loose any meaning
from $Z=64$ since they do not take into account the role played by
the $\pi 0h_{11/2}$ orbital.
For the sake of completeness, in Fig. \ref{tinJ2p} we report the
behavior of the excitation energies of the yrast $J=2^+$ states for
the heavy tin isotopes, but the calculations could not be carried out
beyond $N=90$ owing to the computational complexity.

To summarize, the results of our calculations show that both SP
energies and monopole properties of the orbitals without their own
spin-orbit partner in the chosen model space are quite sensitive to
the cutoff value.
We can then infer that these quantities are sensitive to the missing
3NF too.
Finally, the comparison with the available data shows that, since the
inclusion of the 3NF should eliminate - or at least reduce - the
cutoff dependence and provide a better agreement with experiment,
the larger the cutoff the smaller is the role of the missing 3NF.

\section{Summary}
In this paper we have presented the results of a study of the role of
3NF in realistic shell-model calculations for nuclei beyond
doubly-closed $^{132}$Sn.
In particular, we have focused our attention on the $N=82$ isotonic
and the heavy tin isotopic chains, employing effective shell-model
hamiltonians derived from two low-momentum on-shell-equivalent
potentials characterized by two different cutoffs.

Our investigation of the 3NF relevance has been carried out
indirectly, by comparing the results with the cutoffs $\Lambda=2.1$
and 2.6 fm$^{-1}$ and tracing back the sensitivity to this parameter
to the role of the missing 3NF.

The results show that the orbitals lacking their spin-orbit partner in
the chosen model space are quite affected by the cutoff value, 
a better agreement with experiment being obtained when employing
a larger cutoff.

\end{document}